# Resource Letter GrW-1: Gravitational Waves

Joan M. Centrella

*Laboratory for High Energy Astrophysics, NASA/Goddard Space Flight Center, Greenbelt, Maryland 20771*

This Resource Letter provides a guide to the literature on the physics and astrophysics of gravitational waves. Journals, books, reports, archives, and websites are provided as basic resources and for current research frontiers in detectors, data analysis, and astrophysical source modeling.

## I. INTRODUCTION

The phenomenon of gravitational radiation was one of the first predictions of Einstein's general theory of relativity. Progress in understanding this radiation theoretically was slow at first, owing to the difficulty of the nonlinear field equations and the subtleties of their physical effects. The experimental side of this subject also has taken a long time to develop. Gravitational radiation was detected indirectly in 1974 by J. Taylor and R. Hulse, who observed its effects on the orbital period of a binary system containing two neutron stars, one of them a pulsar (PSR 1913 + 16). Efforts to detect gravitational waves directly have been severely challenged by the extreme weakness of the waves impinging on the Earth. However, as the 21st century begins, observations of the gravitational waves from astrophysical sources such as black holes, neutron stars, and stellar collapse are expected to open a new window on the universe. Vigorous experimental programs centered on ground-based detectors are being carried out worldwide, and a space-based detector is in the planning stages. On the theoretical side, much effort is being expended to produce robust models of the astrophysical sources and accurate calculations of the waveforms they produce. In this Resource Letter, a set of basic references will be presented first, to provide a general introduction to and overview of the literature in this field. The focus then will shift to highlighting key resources in more specialized areas at the forefront of current research.

## II. BASIC RESOURCES

### A. Journals

Most of the important results in gravitational-wave physics and astrophysics are published in the following journals:

*Physical Review D*
*Physical Review Letters*
*Classical and Quantum Gravity*
*Astrophysical Journal*
*Astrophysical Journal Letters*
*Living Reviews in Relativity (electronic journal: www.livingreviews.org)*
*Astronomy and Astrophysics*
*Monthly Notices of the Royal Astronomical Society*
*Review of Scientific Instruments*
*Measurement Science and Technology*
*Journal of Physics E*
*Journal of the Optical Society of America*

*Physics Letters A*
*Optics Letters*

## B. Electronic Archives

Most of the important papers in the areas of gravitational-wave theoretical source prediction, measurement theory, and analysis from the mid-1990s onward are posted on one or both of the following e-print archives:

http://lanl.arxiv.org/archive/gr-qc/ This website (abbreviated gr-qc in the references in Secs. III. C-D below) features papers in the areas of general relativity and quantum cosmology.

http://lanl.arxiv.org/archive/astro-ph/ This website (abbreviated astro-ph in the references in Secs. III. C-D below) has papers in the general category of astrophysics.

The LIGO Document Control Center is a repository of experimental papers, mainly focused on LIGO: http://admdbsrv.ligo.caltech.edu/dcc/. The VIRGO project maintains a bibliography of articles from roughly 1990 – 1999 at the website http://www.pg.infn.it/virgo/local/biblio.htm in the following four categories: gravitational-wave theory and experiment, data analysis in gravitational-wave detectors, internal friction and thermal noise, creep and acoustical emissions.

In addition, the NASA Astrophysics Data System provides, without charge, the ability to search the abstracts and access the full text of many of the relevant papers and journals in which research on gravitational waves is published: see http://adswww.harvard.edu/.

## C. Conference Proceedings

Researchers in gravitational-wave physics and astrophysics depend heavily on conferences and their proceedings for the dissemination of new results. A representative sampling of recent proceedings volumes is given below.

1. **Proceedings of the 4th Edoardo Amaldi Conference on Gravitational Waves**, Classical and Quantum Gravity **19**, 1227 – 2049 (2002). This special issue, available freely at the journal's website (http://www.iop.org/EJ/S/UNREG/nB8zy.1HfQeuoB8FLR2b4A/toc/0264-9381/19/7 ) is edited by David Blair. (A)
2. **Astrophysical Sources for Ground-Based Gravitational Wave Detectors**, AIP Conference Proceedings 575, edited by J. Centrella (AIP, Melville, NY, 2001). (A)
3. **Proceedings of the 3rd International LISA Symposium**, Classical and Quantum Gravity **18**, 3965-4164 (2001). This special issue is edited by Bernard Schutz. (A)
4. **Gravitational Waves: A Challenge to Theoretical Physics**, edited by V. Ferrari, J. Miller, and L. Rezzolla (Abdus Salam International Centre for Theoretical Physics, Trieste, 2001). (A)
5. **Laser Interferometric Space Antenna: Proceedings of the Second International LISA Symposium**, AIP Conference Proceedings 456, edited by W. Folkner (AIP, Woodbury, NY, 1998). (A)
6. **Proceedings of the International Conference on Gravitational Waves: Sources and Detectors**, edited by I. Ciufolini and F. Fidecaro (World Scientific, Singapore, 1997). (A)

**D. Textbooks and Expositions**

   The topic of gravitational waves is usually introduced as part of a course in general relativity. The following is a representative sample of useful textbooks. The first two books are at the level of advanced undergraduates or beginning graduate students. The book by Stephani is at the introductory graduate level.

7.  **A First Course in General Relativity**, B. Schutz (Cambridge University Press, Cambridge, 1985). (I)
8.  **Introducing Einstein's Relativity**, R. d'Inverno (Clarendon Press, Oxford, 1992). (I)
9.  **General Relativity: An Introduction to the Theory of the Gravitational Field**, H. Stephani, 2$^{nd}$ ed. (Cambridge University Press, Cambridge, 1990). (A)

The next four books are at the level of advanced graduate students. Although they were originally published roughly 20 - 30 years ago, they remain standard texts and references for both students and researchers. Readers should be aware that much of the astrophysical material in the applications in these references is now out of date.

10. **Gravitation**, C. Misner, K. Thorne, and J. Wheeler (W. H. Freeman, New York, 1973). (A)
11. **Gravitation and Cosmology: Principles and Applications of the General Theory of Relativity**, S. Weinberg (Wiley, New York, 1972). (A)
12. **Problem Book in Relativity and Gravitation**, A. Lightman, W. Press, R. Price, and S. Teukolksy (Princeton University Press, Princeton, 1975). (A)
13. **General Relativity**, R. Wald (University of Chicago Press, Chicago, 1984). (A)

There also are a number of useful expositions on gravitational waves, at a variety of levels. The first two books are accessible to general readers, and also provide overviews and perspectives of interest to more specialized researchers. The third book presents an introduction to gravitational-wave detectors at the advanced undergraduate level and above, and the fourth at the graduate level.

14. **Einstein's Unfinished Symphony: Listening to the Sounds of Space-Time**, M. Bartusiak (Joseph Henry Press, Washington, D.C., 2000). This well-researched book provides a history of gravitational-wave physics, with interviews with many of the principal players. (E)
15. **Gravity's Fatal Attraction: Black Holes in the Universe**, M. Begelman and M. Rees (Scientific American Library, New York, 1998). This book covers many of the strongest astrophysical sources of gravitational waves. (E)
16. **Fundamentals of Interferometric Gravitational Wave Detectors**, P. Saulson (World Scientific, Singapore, 1994). This book gives a comprehensive introduction to gravitational-wave detectors with the emphasis on laser interferometers. (I)
17. **The Detection of Gravitational Waves**, D. Blair (ed.) (Cambridge University Press, Cambridge, 1991). This book provides an introduction to gravitational waves, with the emphasis on detection. (A)

**E. Websites**

   The AstroGravS website contains a catalog of computed gravitational waveforms that can be downloaded, as well as useful information on astrophysical sources and links to the literature.

18. http://astrogravs.gsfc.nasa.gov

The following websites are maintained by various ground-based gravitational-wave detection efforts; such detectors will be sensitive to high-frequency gravitational waves, roughly 10 – 1000 Hz.

19. http://www.ligo.caltech.edu   This is the website for the US LIGO project.
20. http://www.geo600.uni-hannover.de/   This is the website for the GEO-600 interferometer.
21. http://www.virgo.infn.it/   This is the website for the VIRGO interferometer.
22. http://tamago.mtk.nao.ac.jp/   This is the website for the Japanese TAMA detector.
23. http://www.anu.edu.au/Physics/ACIGA/ This is the website for the Australian Consortium for Interferometric Gravitational Astronomy.
24. http://igec.lnl.infn.it/ This website for the International Gravitational Event Collaboration (IGEC) has links to various resonant mass (or "bar") detector groups.
25. http://www.minigrail.nl/  This is the website for a team developing spherical resonant mass detectors.

These three websites are dedicated to the joint NASA/ESA collaboration on the Laser Interferometric Space Antenna (LISA) effort.  This space-based detector will be sensitive to low-frequency gravitational waves, with frequencies roughly $10^{-4} – 10^{-1}$ Hz.

26. http://lisa.jpl.nasa.gov/
27. http://lisa.gsfc.nasa.gov
28. http://sci.esa.int/home/lisa/

## III.  Current Research Frontiers

Research on gravitational waves is an active and growing field, encompassing a broad range of work in experimental, theoretical, and computational physics.  In this section, several key references in the main areas of activity will be highlighted to provide entry points into the literature.

### A.  General Articles

The following general articles provide useful introductions and surveys of the field of gravitational radiation research.

29. "New physics and astronomy with the new gravitational-wave observatories,"  S. Hughes, *et al.*, astro-ph/0110349 (Proceedings of the 2001 Snowmass Meeting). (I) This is a recent overview at a fairly general level of the growing field of gravitational-wave physics and astrophysics.
30. "Gravitational Radiation and the Validity of General Relativity," C. Will, Phys. Today **52**, 38 – 43 (October 1999). (I)  This article gives an accessible accounts of how observations of gravitational waves can be used to test general relativity.
31. "Gravitational Radiation," K. Thorne, in 300 Years of Gravitation, edited by S. Hawking and W. Israel (Cambridge University Press, Cambridge, 1987),  pp. 330 – 458. (A)  This article, although published 15 years ago, remains an important general introduction to research in gravitational radiation at a research level.
32. "The Detection of Gravitational Waves," J. A. Lobo, gr-qc/0202063 (Proceedings of the ERE-2001 Conference, Madrid, September 2001). (A)  This article reviews the theory of detection by laser interferometers and resonant mass detectors.
33. "Gravitational Wave Detection by Interferometry (Ground and Space)," S. Rowan and J. Hough, Living Reviews in Relativity, http://www.livingreviews.org/Articles/Volume3/2000-3hough/index.html. (A)  This article is a recent review of interferometric detectors at a research level.

34. "LISA Mission Overview," K. Danzmann, for the LISA Study Team, Advances in Space Research, **25**, 1129 – 1136 (2000). (I) This is an accessible introductory article on LISA.
35. "LIGO and the Detection of Gravitational Waves," B. Barish and R. Weiss, Phys. Today **52**, 44 – 50 (October 1999). (I) This is an accessible introduction to gravitational wave detection by LIGO.
36. "LIGO: The Laser Interferometer Gravitational-wave Observatory," A. Abramovici, *et al.*, Science **256**, 325 – 333 (1992). (I) This is an accessible introductory article on LIGO.

## B. Detectors

Much of the current information available on the experimental efforts can be found in articles in the conference proceedings in Sec. II C and at the websites listed in Sec. II E. Some reports of particular interest include the following:

37. http://www.ligo.caltech.edu/~ligo2/ This website and the documents available there detail the plans for the advanced LIGO interferometers.
38. http://lisa.jpl.nasa.gov/documents/ppa2-09.pdf The Pre-Phase A Report, second edition (July 1998), for the LISA project is available here.

In addition, the following journal and proceedings articles provide a representative sampling of the literature on detectors.

**Ground-based interferometers:**
39. "Lock acquisition of a gravitational wave interferometer," M. Evans, *et al.*, accepted for publication in Optics Letters (2002).
40. "Readout and control of a power-recycled gravitational-wave antenna," P. Fritschel, *et al.*, Appl. Optics **40**, 4988 – 4998 (2001).
41. "Conversion of conventional gravitational-wave interferometers into QND interferometers by modifying their input and/or output optics," H. Kimble, *et al.*, Phys. Rev. D **65**, 022002 (2002).
42. "Seismic Isolation for Advanced LIGO," R. Abbott, *et al.*, Class. Quantum Grav. **19**, 1591-1597 (2002).
43. "Signal recycled laser-interferometer gravitational-wave detectors as optical springs," A. Buonanno and Y. Chen, Phys. Rev. D **65**, 042001 (2002).
44. "Fused Silica Suspensions for Advanced Gravitational Wave Detectors," S. Rowan, *et al.*, in Gravitational Wave Detection II: Proceedings of the 2<sup>nd</sup> TAMA International Workshop on Gravitational Wave Detection 1999 (Universal Academic Press, Tokyo, 2000), pp. 203-215.
45. "Thermodynamical fluctuations and photo-thermal shot noise in gravitational wave antennae," V. Braginsky, M. Gorodetsky, and S. Vyatchanin, Phys. Lett. A **264**, 1-10 (1999).
46. "Internal thermal noise in the LIGO test masses: a direct approach," Y. Levin, Phys. Rev. D **57**, 659-663 (1998).
47. "Cryogenic cooling of a sapphire mirror-suspension for interferometric gravitational wave detectors," T. Uchiyama, *et al.*, Phys. Lett. A **242**, 211-214 (1998).
48. "Principles of Calculating Alignment Signals in Complex Resonant Optical Interferometers," Y. Hefetz, N. Mavalvala, and D. Sigg, Opt. Soc. Am. B **14**, 1597-1605 (1997).
49. "How an interferometer extracts and amplifies power from a gravitational wave," P. Saulson, Class. Quant. Grav. **14**, 2435-2454 (1997).

50. "Optical Bars in Gravitational Wave Antennas," V.B. Braginsky, M.L. Gorodetsky, F.Ya. Khalili, Phys. Lett. A **232** 340-348 (1997).
51. "Suspension of Detection Masses for the Virgo Interferometer," F. Fidecaro, *et al.*, Nuc. Phys. B (Proc. Suppl.) **48**, 110-112 (1996).
52. "Recycling in laser-interferometric gravitational-wave detectors," B. Meers, Phys. Rev. D **38**, 2317–2326 (1988).

**Resonant mass detectors:**
53. "Wideband Dual Sphere Detector of Gravitational Waves," M. Cerdonio, *et al.*, Phys. Rev. Lett. **87**, 031101 (2001).
54. "Initial operation of the International Gravitational Event Collaboration," G. Prodi, *et al.*, Int. J. Mod. Phys. D **9**, 237-245 (2000).
55. "Search for gravitational radiation with the Allegro and Explorer detectors," P. Astone, *et al*., Phys. Rev. D **59**, 122001 (1999).
56. "Calibration and sensitivity of resonant-mass gravitational wave detectors," A. Morse, *et al*., Phys. Rev. D **59**, 062002 (1999).
57. "Gravitational-wave stochastic background detection with resonant-mass detectors," S. Vitale, *et al.*, Phys. Rev. D **55**, 1741–1751 (1997).
58. "The Allegro gravitational wave detector: Data acquisition and analysis," E. Mauceli, *et al*., Phys. Rev. D **54**, 1264-1275 (1996).
59. "Spherical Gravitational Wave Antennas and the Truncated Icosahedral Arrangement," S. Merkowitz and W. Johnson, Phys. Rev. D **51**, 2546 (1995).

**Space-based interferometers:**
60. "Improving the Sensitivity of LISA," K. Nayak, *et al.*, gr-qc/0210014 (submitted to Classical and Quantum Gravity).
61. "The LISA Optimal Sensitivity," T. Prince, *et al.*, gr-qc/0209039 (submitted to Phys. Rev. D).
62. "LISA Interferometer sensitivity to spacecraft motion," M. Petersein, *et al.*, Advances in Space Research, **25**, 1143 – 1147 (2000).
63. **"**Laser Frequency Stabilisation For LISA: Experimental Progress," P.W. McNamara, H. Ward, J. Hough, Advances in Space Research, **25**, 1137 – 1142 (2000).
64. "Angular Resolution of the LISA Gravitational Wave Detector," C. Cutler, Phys. Rev. D **57** 7089-7102 (1998).

## C. Astrophysical Sources of Gravitational Waves

The expected sources of gravitational radiation are all astrophysical in nature and comprise some of the most interesting and exotic objects in the universe, such as black holes. The following pair of companion articles presents connections between black holes and gravitational waves, the first from the viewpoint of fundamental physics and the second from astrophysics:

65. "Probing Black Holes and Relativistic Stars with Gravitational Waves," K. Thorne, in <u>Black Holes and Relativistic Stars</u>, edited by R. Wald (University of Chicago Press, Chicago, 1998), pp. 41 – 78. (A)
66. "Astrophysical Evidence for Black Holes," M. Rees, in <u>Black Holes and Relativistic Stars</u>, edited by R. Wald (University of Chicago Press, Chicago, 1998), pp. 79 – 102. (A)

Inspiralling compact binaries constitute one of the most important classes of gravitational-wave sources. When the binary components are neutron stars and/or stellar black holes, the gravitational radiation emitted towards the end of the inspiral is in the high frequency range of roughly 10 – 1000 Hz, making the systems targets for ground-based detectors. If the binary components are massive black holes, with masses in the range of roughly $10^4 - 10^7$ times the mass of the sun, the gravitational waves are of low frequency, roughly $10^{-4} - 10^{-1}$ Hz, and may be detected by the space-based LISA.

During the early stages of the inspiral, the binary components are widely separated and can be treated as point particles. Post-Newtonian methods have been used to calculate the gravitational waves in this regime for many years; more recently, an effective one-body approach has been used. Some representative papers include:

67. "Post-Newtonian gravitational radiation and equations of motion via direct integration of the relaxed Einstein equations. II. Two-body equations of motion to second post-Newtonian order, and radiation-reaction to 3.5 post-Newtonian order," M. Pati and C. Will, Phys. Rev. D **65**, 104008 (2002). (A)

68. "Gravitational Radiation from Post-Newtonian Sources and Inspiralling Compact Binaries," L. Blanchet, Living Reviews in Relativity, http://www.livingreviews.org/Articles/Volume5/2002-3blanchet/index.html. (A)

69. "Coalescence of Two Spinning Black Holes: An Effective One-Body Approach," T. Damour, Phys. Rev. D **64,** 124013 (2001). (A)

70. "Gravitational waveforms from inspiralling compact binaries to second-post-Newtonian order," L. Blanchet, B. Iyer, C. Will, and A. Wiseman, Class.Quant.Grav. **13,** 575-584 (1996). (A)

Once the binary components get close to merger, full numerical simulations are needed to calculate the waveforms. The following article is a recent review of current efforts to simulate compact binary mergers numerically:

71. "Numerical Relativity and Compact Binaries," T. Baumgarte and S. Shapiro, gr-qc/0211028 (to appear in Physics Reports).

For binary black-hole systems which have no gas or other matter, the waveforms from a calculation can be scaled with the masses and spins of the components to apply equally to stellar black hole systems (producing high-frequency gravitational waves) or massive black hole systems (producing low frequency waves). The following papers highlight some of the recent successes and challenges in the simulation of binary black hole mergers:

72. "Modeling gravitational radiation from coalescing binary black holes," J. Baker, M. Campanelli, C. Lousto, and R. Takahashi, Phys. Rev. D **65**, 124012 (2002). (A)

73. "Grazing Collisions of Black Holes via the Excision of Singularities,", S. Brandt, *et al.*, Phys. Rev. Lett. **85**, 5496-5499 (2000). (A)

74. "The 3D Grazing Collision of Two Black Holes," M. Alcubierre, *et al.*, Phys. Rev. Lett. **87**, 271103 (2001) (A)

If one or more of the binary components are neutron stars, the equations of hydrodynamics also must be solved. Representative papers in current research include:

75. "Post-Newtonian SPH calculations of binary neutron star coalescence. III. Irrotational systems and gravitational wave spectra," J. Faber and F. Rasio, Phys. Rev. D **65**, 084042 (2002). (A)

76. "Simulation of merging binary neutron stars in full general relativity: Γ =2 case," M. Shibata and K. Uryu, Phys. Rev. D **61**, 064001 (2000). (A)

Compact binaries within the Milky Way Galaxy are an important source of low-frequency gravitational waves for LISA. While some of these galactic binaries should be detectable individually, the gravitational waves from the vast number of such binaries will produce a stochastic background of "noise" for LISA. The following papers discuss this class of sources:

77. "The gravitational wave signal from the Galactic disk population of binaries containing two compact objects," G. Nelemans, L. Yungelson, and S. Portegies Zwart, astro-ph/0105221, Astron. Astroph., in press. (A)

78. "Low Frequency Gravitational Waves from White Dwarf MACHO Binaries," W. Hiscosk, *et al.*, Astrophys.J. **540**, L5-L8 (2000). (A)

An "extreme-mass-ratio" binary, consisting of a neutron star or stellar black hole inspiralling into a massive black hole at the center of a galaxy, is one of the most interesting sources for LISA. The gravitational radiation produced as the smaller object inspirals into the massive black hole can provide a map of the gravitational potential of this exotic central object. This situation is explored in the following articles:

79. "Evolution of circular, non-equatorial orbits of Kerr black holes due to gravitational-wave emission: II. Inspiral trajectories and gravitational waveforms," S. Hughes, Phys. Rev. D **64**, 064004 (2001). (A)

80. "Gravitational Waves from a Compact Star in a Circular, Inspiral Orbit, in the Equatorial Plane of a Massive, Spinning Black Hole, as Observed by LISA," L. S. Finn and K. Thorne, Phys. Rev. D **62**, 124021 (2000). (A)

There is a variety of other sources of gravitational radiation, including the gravitational collapse of stars and supermassive stars, dynamical "bar" instabilities in rapidly rotating stars, crustal deformations of neutron stars, and cosmological graviational waves. The following list of papers gives a sample of recent work in some of these areas:

81. "Gravitational Wave Emission From Core-Collapse of Massive Stars," C. Fryer, D. Holz, and S. Hughes, Astrophys. J. **565** 430-446 (2002). (A)

82. "Gravitational waves from relativistic rotational core collapse," H. Dimmelmeier, J. Font, and E. Mueller, Astrophys. J. **560**, L163-L166 (2001). (A)

83. "Dynamical Bar Instability in Rotating Stars: Effect of General Relativity," M. Saijo, *et al.*, Astrophys J. **548**, 919-931 (2001). (A)

84. "Gravitational Waves from Long-Duration Simulations of the Dynamical Bar Instability," K. New, J. Centrella, and J. Tohline, Phys. Rev. D **62**, 064019 (2000). (A)

85. "Collapse of a Rotating Supermassive Star to a Supermassive Black Hole: Post-Newtonian Simulations," M. Saijo, *et al.*, Astrophys. J. **569**, 349-361 (2002). (A)

86. "Evolution of Differentially Rotating Supermassive Stars to the Onset of Bar Instability," K. New and S. Shapiro, Astrophys. J. **548**, 439-446 (2001). (A)

## D. Gravitational-Wave Data Analysis

The field of gravitational-wave data analysis, although in its infancy, is developing rapidly to meet the needs of the experimental efforts. The resonant-mass detector community has become organized and has started exchanging data. The kilometer-scale ground-based interferometers are just beginning to take data. And the consideration of data analysis strategies is playing an important role in the design of LISA.

The conference proceedings listed in Sec. IIC above are a good source of articles on current gravitational-wave data analysis efforts. In addition, the following journal articles provide a representative sampling of the current literature:

87. "Data analysis of gravitational-wave signals from spinning neutron stars. IV. An all-sky search," P. Astone, *et al.*, Phys. Rev. D **65**, 042003 (2002). (A)

88. "Robust statistics for deterministic and stochastic gravitational waves in non-Gaussian noise I: Frequentist analyses," B. Allen, *et al.*, Phys. Rev. D **65**, 122002 (2002). (A)

89. "Robust statistics for deterministic and stochastic gravitational waves in non-Gaussian noise I: Bayesian analyses," B. Allen, *et al.*, gr-qc/0205015. (A)

90. "A data-analysis strategy for detecting gravitational-wave signals from inspiraling compact binaries with a network of laser-interferometric detectors," A. Pai, S. Shurandar, and S. Bose, Phys. Rev. D **64**, 042004 (2001). (A)

91. "An excess power statistic for detection of burst sources of gravitational radiation," W. Anderson, *et al.*, Phys. Rev. D **63**, 042003 (2001). (A)

92. "An efficient filter for detecting gravitational wave bursts in interferometric detectors," T. Pradier, *et al.*, Phys. Rev. D **63**, 042002 (2001). (A)

93. "Mapping the gravitational wave background," N. Cornish, Class. Quant. Grav. **18**, 4277-4292 (2001).

94. "Search techniques for gravitational waves from black-hole ringdowns," J. Creighton, Phys. Rev. D **60**, 022001 (1999). (A)

95. "Time-frequency detection of Gravitational Waves," W. Anderson and R. Balasubramanian, Phys. Rev. D **60**, 102001 (1999). (A)

96. "Measuring gravitational waves from binary black hole coalescences: II. The waves' information and its extraction, with and without templates," E. Flanagan and S. Hughes, Phys. Rev. D **57**, 4566-4587 (1998). (A)

97. "Measuring gravitational waves from binary black hole coalescences: I. Signal to noise for inspiral, merger, and ringdown," E. Flanagan and S. Hughes, Phys. Rev. D **57**, 4535-4565 (1998). (A)